\documentstyle[twocolumn,aps,prl,epsfig,floats]{revtex}  

\newcommand{\cbcex}{\langle\overline{\chi}\chi\rangle}
\newcommand{\chup}{$\chi U\phi$}

\newcommand{\eqref}[1]{(\ref{#1})}

\newcommand\fdfig[1]{%
  \psfig{file=#1,angle=90,width=\hsize,bbllx=65,bblly=15,bburx=540,bbury=780}}%

\begin{document}
\draft
\title{ \hfill {\small PITHA 97/42, HLRZ1997\_65}\\
       Strongly coupled U(1) lattice gauge theory \\
       as a microscopic model of Yukawa theory
}
\author{W.~Franzki and J.~Jers{\'a}k}
\address{Institut f{\"u}r
Theoretische Physik E, RWTH Aachen, Germany}
\date{\today}
\maketitle
\begin{abstract}
  Dynamical chiral symmetry breaking in a strongly coupled U(1) lattice gauge
  model with charged fermions and scalar is investigated by numerical
  simulation. Several composite neutral states are observed, in particular a
  massive fermion. In the vicinity of the tricritical point of this model we
  study the effective Yukawa coupling between this fermion and the Goldstone
  boson. The perturbative triviality bound of Yukawa models is nearly
  saturated. The theory is quite similar to strongly coupled Yukawa models for
  sufficiently large coupling except the occurrence of an additional state --
  a gauge ball of mass about half the mass of the fermion.
\end{abstract}
\vspace{1cm}
\narrowtext 

The question whether the Higgs-Yukawa mechanism of symmetry breaking
and particle mass generation in the contemporary high energy physics
might be an effective theory to some more fundamental renormalizable
quantum field theory with dynamical symmetry breaking has been raised
many times. It is of particular interest with respect to the large top
quark mass, stimulating e.\,g.\ the idea of topcolor. The pursuit of
this question leads frequently to devising strongly coupled gauge
theories beyond the standard model with speculative nonperturbative
dynamics.  It might be inspiring to have prototypes of such
``microscopic'' gauge models of Yukawa theory with a solid and
accessible dynamical framework. Numerical lattice methods are
justified for this purpose provided a lattice model can be found with
required basic dynamical properties, though neglecting various
phenomenological aspects. In particular, these properties should be
preserved when approaching the continuum limit, i.e. the model should
be nonperturbatively renormalizable.

A promising lattice model has been suggested in Ref.~\cite{FrJe95a}.
This ``\chup\ model'' consists of a charged fermion field $\chi$ with
strong vectorlike coupling to compact U(1) gauge field $U$, and a
scalar field $\phi$ of the same charge.  Here we summarize the results
of our extensive systematic investigations of this model in four
dimensions by means of numerical simulations with dynamical fermions.
Though we cannot decide by currently available means whether the model
is renormalizable, we find some encouraging properties. The detailed
account is given in a parallel paper \cite{FrJe98b} and in some
previous works \cite{FrFr95,FrFr96,Fr97a,Fr97b}.

The same model has been investigated also in lower dimensions. There is little
doubt that the \chup\ model defines a renormalizable quantum field theory in
two dimensions, belonging to the universality class of the chiral Gross-Neveu
model \cite{FrJe96c}. Also first results in three dimensions suggest
renormalizability \cite{BaFo98}.

Finding a ``promising'' lattice model meets two major difficulties.  Strongly
coupled lattice gauge theories with fermions exhibit frequently the required
dynamical chiral symmetry breaking. However, this phenomenon comes mostly
along with undesirable confinement of fermions acquiring mass.  Furthermore,
at strong coupling, it is difficult to find a second order phase transition
required for an approach to continuum.

In the \chup\ model, the scalar field $\phi$ helps to solve both problems.
Firstly, it shields the fermion charge and gives rise to an unconfined, i.e.
physical massive fermion $F = \phi^\dagger\chi$ in the phase with chiral
symmetry broken dynamically by the gauge interaction (Nambu phase).  Secondly,
the scalar {\em suppresses} this symmetry breaking and at sufficiently strong
gauge coupling induces a second order transition to a chiral symmetric (Higgs)
phase, thus opening a way to continuum.

We find that one particular point of the phase transition line, the
tricritical point E, represents a new kind (new universality class) of
dynamical symmetry breaking mechanism in strongly coupled gauge
theories with fermions in four dimensions.  We present some evidence
that a strongly coupled effective Yukawa theory results. It describes
the interaction between the composite fermion $F$ and some
$\overline{\chi} \chi$ ``mesons''.  At this point the model could
serve as a microscopic model of Yukawa theory. But in addition, the
mechanism produces also a scalar gauge ball.

To explain these findings, we now summarize the most relevant features of the
model (more details can be found in \cite{FrJe98b}). The model is defined by
the action
\begin{eqnarray*}
  S &=& S_\chi + S_U + S_\phi \; , \\ 
  S_\chi & = & \frac{1}{2} \sum_x
   \overline{\chi}_x \sum_{\mu=1}^4 \eta_{x\mu} (U_{x,\mu} \chi_{x+\mu} -
   U^\dagger_{x-\mu,\mu} \chi_{x-\mu}) \\ 
         & &+{am_0} \sum_x \overline{\chi}_x\chi_x\; ,\\ 
  S_U & = & -\beta \sum_P \cos(\Theta_P)\; , \\ 
  S_\phi & = & - {\kappa}
   \sum_x \sum_{\mu=1}^4 (\phi^\dagger_x U_{x,\mu} \phi_{x+\mu} + h.c.).
\end{eqnarray*}
Here $\Theta_P \in [0,2\pi)$ is the plaquette angle, i.e. the argument
of the product of U(1) gauge field link variables $U_{x,\mu}$ along a
plaquette $P$.  Taking $\Theta_P = a^2gF_{\mu\nu}$, where $a$ is the
lattice spacing, and $\beta = 1/g^2$, one obtains for weak coupling
$g$ the continuum gauge action $S_U=\frac{1}{4} \int
d^4xF_{\mu\nu}^2$. The staggered fermion field $\chi$ has (real) bare
mass $am_0$ in lattice units. It leads to four fermion species in the
continuum limit. Sign factors $\eta_{x\mu}$ are standard for staggered
fermions. The complex scalar field is constrained, $|\phi| = 1$. As
its ``hopping parameter'' $\kappa$ increases, it drives the model into
the usual Higgs phase.

We note that there is no Yukawa coupling between $\chi$ and $\phi$, as both
fields have the same charge. The model has U(1) global chiral symmetry in the
limit of vanishing fermion bare mass in physical units, $m_0= 0$. This is the
case we are really interested in. The numerical simulations have to be
carried out at nonvanishing $m_0$, however, and an extrapolation to the chiral
limit performed. For $am_0 \ne 0$, the chiral transformation relates the model
at $\pm am_0$.

\begin{figure}
  \begin{center}
    \fdfig{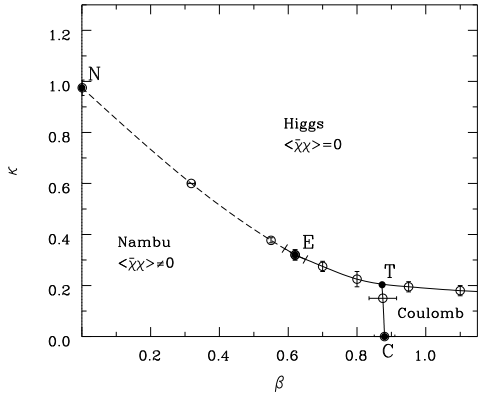}%
    \caption[xxx]{%
      Phase diagram of the \chup\ model in the chiral limit, $m_0 =
      0$. The $\beta = 0$ limit corresponds to the NJL model. The NE
      line is a line of second order phase transitions. Other lines
      are lines of first order transitions. In the Nambu phase, chiral
      symmetry is broken dynamically and the fermion $F =
      \phi^\dagger\chi$ is massive. In the Higgs phase, $F$ is
      massless. E is the tricritical point where the scaling behavior
      is different from the NJL model. The Coulomb and Higgs phases
      extend until $\beta = \infty$. The data points are positions of
      the phase transitions extrapolated to the infinite volume.}
    \label{fig:pd4d2}
  \end{center}
\end{figure}%

The phase diagram at $m_0= 0$ is shown in Fig.~\ref{fig:pd4d2}. One limit case
is $\kappa = 0$, corresponding to the massless compact lattice QED. Another
important limit case is $\beta = 0$. Here the model can be rewritten exactly
as the lattice Nambu--Jona-Lasinio (NJL) model \cite{LeShr87a} with the
critical point N. The strong four-fermion coupling of that model corresponds
to small $\kappa$. The Nambu phase thus connects the broken chiral symmetry
phases of both these limit models. Only a part of its boundary, the NE line,
represents second order phase transitions.

Point E is far away from any limit case and it does not appear to be
accessible by any reliable analytic method, neither on the lattice nor in
continuum. It is ``tricritical'' because in the full parameter space
(including $am_0$) there are, apart from NE, two further second order ``wing''
lines entering E from the positive and negative $am_0$ directions. The
existence of a common point E of these three second order lines is neither
predicted nor understood. The evidence is purely numerical, but quite strong
\cite{FrJe98b}. Its position is $\beta_{\rm E} = 0.62(3), \kappa_{\rm E} =
0.32(2)$.

The importance of the point E roots in the experience from statistical
mechanics that a tricritical point belongs to a universality class
different from that of any of the second order lines entering into
it\cite{LaSa84}.  (Basic properties of tricritical points relevant in
our context are summarized e.g.  in \cite{EvJe85}.) The usual
universality of second order lines suggests, and this is supported by
our earlier investigations \cite{FrFr95}, that the whole NE line
except the point E corresponds to the same continuum model as the
point N, the NJL model. The gauge field is presumably auxiliary and
the model is therefore of limited interest there. However, the point E
is expected to be different, gauge field playing an important role.

To verify this expectation, we have investigated critical exponents
and spectrum of the model in the vicinity of E. The study of exponents
uses advanced techniques of statistical physics and is described in
\cite{FrJe98b}.  Here we only mention the found value $\nu_t \simeq
1/3$ of the correlation length tricritical exponent. It is different
from $\nu \simeq 1/2$ along the NE line confirming the particularity
of E. It also differs from the prediction of the classical theory of
tricritical points usually believed to hold in four dimensions
\cite{LaSa84} and predicting $\nu_t = 1/2$. This indicates that the
point E is a tricritical point with important role of quantum
fluctuations.

Some insight into the continuum physics, which might be obtained at the point
E, is provided by the spectrum and its scaling behavior. The masses $am_{\rm
  Q}$ are in lattice units and thus only their ratios are physical. They are
determined in the Nambu phase without any gauge fixing from the correlation
functions of various gauge invariant composite operators ${\rm Q}$. In this
sense the massive physical fermion $F$, as well as other physical states, are
composite.  The interaction between them is due to the Van der Waals
remnant of the fundamental interactions.

The fermion mass $am_F > 0$ decreases when the NE line is approached
from the Nambu phase and there it is rather insensitive to the lattice
size. It is small in the Higgs phase on finite lattices, decreasing
with increasing lattice size. The data are consistent with expected
vanishing of $am_{\rm F}$ in the Higgs phase in the infinite volume
limit.

We find several $\overline{\chi}$, $\chi$ bound states and borrow their names
from QCD. One of them is the obligatory pseudoscalar Goldstone boson $\pi$
with the dependence on $am_0$ as required by current algebra. The $\pi$-meson
is massless in the chiral limit in the whole Nambu phase.  It would get
``eaten'' if the global chiral symmetry of the \chup\ model were gauged. This
process would be identical with the standard Higgs mechanism. Thus we do not
need to discuss it and mention it only for completeness.

We obtain some results for the pion decay constant. The data suggest a large
ratio $f_\pi/m_F$ increasing when the point E is approached. However, the
value is sensitive to the lattice volume and we cannot yet extrapolate it to
the infinite volume and continuum limit. If the ratio diverges, it
would indicate the triviality of the model \cite{HaKo97}. Its current
value, $f_\pi/m_F \simeq 1/3$ for $am_F \simeq 0.4$ can be considered
as a lower bound.

Also the $\sigma$ meson in the antifermion-fermion channel is observed. The
$\sigma$ mass is quite dependent both on the lattice size and bare fermion
mass, preventing its prediction in the continuum limit with $m_0=0$. Strong
lattice size dependence of $f_\pi$ and $\sigma$ mass has been observed and
explained by means of the Schwinger-Dyson equations in the limit case $\beta =
0$, the NJL model, and holds apparently also in the vicinity of E.

The $\rho$ mass is insensitive to the lattice size and scales like the fermion
mass, with the approximate value $m_\rho \simeq 2 m_F$. We cannot distinguish
between a bound state and a resonance.

Also a neutral scalar ($S$-boson) is seen appearing as a composite of
$\phi^\dagger$ and $\phi$ and as a state of pure gauge field. In the Higgs
phase it corresponds to the Higgs boson associated with the perturbative Higgs
mechanism occuring in that phase at large $\beta$. In the Nambu phase it is
more natural to interpret the $S$-boson as a scalar gauge ball. The transition
between both interpretations is smooth, however, and the corresponding
channels appreciably mix in the vicinity of E.

The mass $am_S$ is nonzero in all phases and goes to zero when the two
wing critical lines are approached for any $am_0$.  For $am_0 = 0$,
the mass $am_S$ goes to zero only at the point E. This illustrates the
particular character of the tricritical point E. Whereas only one of
the masses $am_F$ or $am_S$ vanishes on each of the second order
lines, they both vanish at the tricritical point. As their ratio
remains finite, both corresponding states are present in the continuum
limit taken at this point. This is the best evidence that the physical
content (universality class) of the point E differs from that of any
of the adjacent second-order lines and thus does not correspond to the
NJL model obtained at N.

\begin{figure}
  \begin{center}
    \fdfig{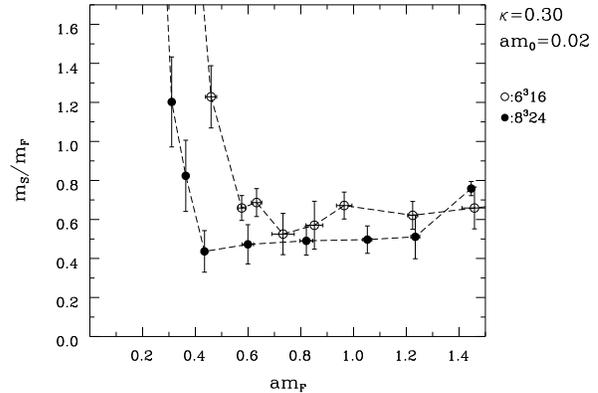}
    \caption{%
      Mass ratio $m_S/m_F$ as function of the fermion mass when point E is
      approached. The data have been obtained on $6^3 16$ (empty circles) and
      $8^3 24$ (full circles) lattices.}
    \label{fig:scalbos}
  \end{center}
\end{figure}

As $am_S$ is only moderately dependent on the lattice size, we can estimate
its scaling behavior when point E is approached. This is shown in
Fig.~\ref{fig:scalbos} for $\kappa = 0.30$, which is approximately the
$\kappa$-coordinate of E. The ratio $m_S/m_F$ remains constant when $am_F$
decreases, as long as finite size effects are negligible.  The sudden rise of
the data at small $am_F$ is due to the strong finite size dependence of $am_F$
in the Higgs phase and shifts correspondingly to smaller $am_F$ when the
lattice volume increases. These observations suggest that the value $m_S/m_F
\simeq 1/2$ would be obtained in the continuum limit at E.

In the rest of this letter we concentrate on the renormalized Yukawa coupling
$y_{\rm R}$ between $F$ and $\pi$. It is obtained from the three-point
function of the corresponding composite operators and thus can be interpreted
as an effective Yukawa coupling, which would describe the interaction between
$F$ and $\pi$ in the regime where their composite structure can be neglected.
Our measurement, performed in a similar way as in \cite{GoHo95}, is described in
detail in \cite{FrJe98b,Fr97b}. In spite of the complexity of the
corresponding expressions, $y_{\rm R}$ is measurable with good precision. The
reason is the strict locality of the used operators. An attempt to measure
also an analogous coupling of the $S$-boson failed because $S$ is described by
extended operators.

\begin{figure}
  \begin{center}
    \fdfig{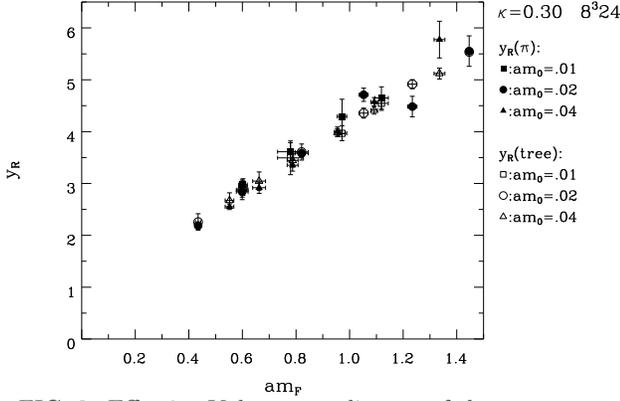}
    \caption{%
      Effective Yukawa coupling $y_{\rm R}$ of the $\pi$ meson to $F$ (full
      symbols), and the tree level relation (\protect\ref{yrtree}) (open
      symbols) as function of $am_F$ for different $am_0$ at $\kappa=0.30$ on
      $8^324$ lattice.}
    \label{fig:yrtree}
  \end{center}
\end{figure}

The renormalized Yukawa coupling $y_{\rm R}$ is presented in
Fig.~\ref{fig:yrtree} (full symbols) for three different $am_0$ at $\kappa =
0.30$ in the Nambu phase close to E. We show it in dependence on $am_F$ and
compare it with the tree level relation
\begin{equation}
  \label{yrtree}
  y_{\rm R}^{\rm (tree)} = \frac{am_F}{\cbcex}\sqrt{Z_\pi}\;.
\end{equation}
The agreement is so good that the open symbols representing (\ref{yrtree}) in
fig.~\ref{fig:yrtree} are nearly invisible.

\begin{figure}
  \begin{center}
    \fdfig{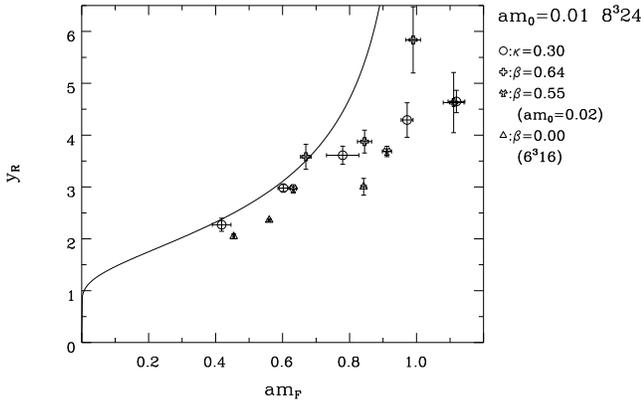}
    \caption{%
      $y_{\rm R}$ of the $\pi$ meson for different couplings along the line
      NE. For $\beta=0.55$ the tree level definition was used. Most data have
      been taken on a $8^3 24$ lattice (at $\beta=0$ on $6^3 16$) with
      $am_0=0.01$ (at $\beta=0.55$ is $am_0=0.02$).  Shown is also the curve
      of maximal $y_{\rm R}$ as it results from the perturbative expansion
      first order in a coresponding Yukawa model. This curve goes to 0 for
      $am_F\rightarrow0$, as expected from triviality.}
    \label{fig:extra_yr}
  \end{center}
\end{figure}%

The values for $y_{\rm R}$ are not significantly dependent on $am_0$.
Therefore we expect that the value for $y_{\rm R}$ in the chiral limit is
rather well represented by our results at $am_0 = 0.01$. In
Fig.~\ref{fig:extra_yr} we show these results for substantially different
gauge couplings. The results close to the point E, for $\beta=0.64$,
$\kappa=0.30$, and $\beta=0.55$, are very similar.  The values of $y_{\rm R}$
at $\beta=0$ (NJL model) are significantly below the other data, however. This
implies that the Yukawa coupling gets stronger as one moves from the NJL model
to the vicinity of E by increasing $\beta$ at fixed $am_F$.



In the interval $y_{\rm R} \simeq 2 - 5 $, where the Yukawa coupling is
determined reliably, we compare our result with the curve of maximal
renormalized coupling, resulting from the first order perturbative calculation
in the Yukawa model (triviality bound):
\begin{equation}
  y_{\rm R\,max} = \frac{1}{\sqrt{2\beta_0|\ln(am_F)|}}\;,\quad \beta_0 =
  \frac{N_F}{4\pi^2}\;,
   \label{up_bound}
\end{equation}
with $N_F=4$, the number of fermions including lattice doublers. In the
lattice Yukawa models this upper bound is nearly saturated by the values of
$y_{\rm R}$ obtained at the maximal possible bare Yukawa coupling
\cite{DeJe92,BoFr93a}. As seen in fig.~\ref{fig:extra_yr}, also in our model
close to E, the data are only slightly below the curve (\ref{up_bound}). Thus
in this interval, $y_{\rm R}$ has the same form as that in the strongly
coupled Yukawa models.

For $y_{\rm R} < 2$, the data are significantly influenced by the
finite size effects, and we cannot make any conclusion. It might be
that some deviation from the Yukawa theory occurs there. The fact that
$y_{\rm R}$ is stronger in the vicinity of E than in the NJL model
suggests the question whether it vanishes in the continuum limit,
$am_F=0$, taken there. Is it zero as in Yukawa theory, or could
triviality be avoided at the tricritical point? A reliable
extrapolation of $y_{\rm R}$ to $am_F=0$ at the tricritical point of
the \chup\ model is a challenge for future investigations.

\vspace{1cm}


We thank D.~Kominis, M.~Lindner, and E.~Seiler for discussions. The
computations have been performed on the Fujitsu S600, VPP500 and VPP300 at
RWTH Aachen, and on the CRAY-YMP and T90 of HLRZ J\"ulich. We thank HLRZ
J\"ulich for hospitality.  The work was supported by DFG.


\bibliographystyle{wunsnot}   

\begin{thebibliography}{10}

\bibitem{FrJe95a}
C.~Frick{ and }\relax J.~Jers\'ak, {\em Phys. Rev.\/} {\bf D52}
  (1995) 340.

\bibitem{FrJe98b}
W.~Franzki{ and }\relax J.~Jers\'ak, {\em Dynamical fermion mass generation at
  a tricritical point in strongly coupled U(1) lattice gauge theory\/}, 
  PITHA 97/43, HLRZ1997\_66, 
  hep-lat/9711039.

\bibitem{FrFr95}
W.~Franzki, C.~Frick, J.~Jers\'ak{,}{ and }\relax X.~Q. Luo,
  {\em Nucl. Phys.\/} {\bf B453} (1995) 355.

\bibitem{FrFr96}
W.~Franzki, C.~Frick, J.~Jers\'ak{,}{ and }\relax X.~Q. Luo,
  {\em Progr. Theor. Phys. Suppl.\/} {\bf 122} (1996) 171.

\bibitem{Fr97a}
W.~Franzki, {\em Nucl. Phys. B (Proc. Suppl.)\/} {\bf 53}
  (1997) 702.

\bibitem{Fr97b}
W.~Franzki, {\em Dynamische Erzeugung von Fermionmassen in stark gekoppelten Eichtheorien auf dem Gitter\/}, Dissertation (RWTH Aachen, 1997).

\bibitem{FrJe96c}
W.~Franzki, J.~Jers\'ak{,}{ and }\relax R.~Welters,
  {\em Phys. Rev.\/} {\bf D54} (1996) 7741.

\bibitem{BaFo98}
I.~M. Barbour, E.~Focht, W.~Franzki, J.~Jers\'ak{,}{ and }\relax
  N.~Psycharis, in preparation.

\bibitem{LeShr87a}
I.-H. Lee{ and }\relax R.~E. Shrock, {\em Phys. Rev. Lett.\/}
  {\bf 59} (1987) 14.

\bibitem{LaSa84}
I.~D. Lawrie{ and }\relax S.~Sarbach, {\em {in }\relax \/}C.~Domb{ and }\relax J.~L. Lebowitz{, eds.}, {\em Phase transitions and critical phenomena\/}, {vol.}~9, {p.}~1 (Acad.
  Press, New York, 1984).

\bibitem{EvJe85}
H.~G. Evertz, J.~Jers\'ak, T.~Neuhaus{,}{ and }\relax P.~M.
  Zerwas, {\em Nucl. Phys.\/} {\bf B251 [FS13]} (1985) 279.

\bibitem{HaKo97}
S.~Hands{ and }\relax J.~B. Kogut, {\em Logarithmic corrections to the equation of state in the SU(2)$\times$SU(2) NJL model\/}, hep-lat/9705015.

\bibitem{GoHo95}
M.~G{\"o}ckeler, R.~Horsley, P.~E.~L. Rakow{,}{ and }\relax
  G.~Schierholz, {\em Phys. Lett.\/} {\bf 353B} (1995) 100.

\bibitem{DeJe92}
A.~K. De{ and }\relax J.~Jers\'ak, {\em {in }\relax \/}A.~J.
  Buras{ and }\relax M.~Lindner{, eds.}, {\em Heavy Flavors\/},
  {p.} 732 (World Scientific, Singapore, 1992).

\bibitem{BoFr93a}
W.~Bock, C.~Frick, J.~Smit{,}{ and }\relax J.~C. Vink,
  {\em Nucl. Phys.\/} {\bf B400} (1993) 309.

\end{thebibliography}


\end{document}